\documentclass[11pt]{article}
\usepackage{color}
\usepackage{amssymb}
\usepackage{amsmath}
\usepackage{subeqnarray}

\usepackage{ifthen}

\usepackage{epsf}
\makeatletter
\@addtoreset{equation}{section}
\makeatother

\def\>{\rangle} 

\def\<{\langle} 

\newcommand{\ams}{${}^1${\it Institute for Theoretical Physics\\
University of Amsterdam, \\
Science Park 904, Postbus 94485, 1090 GL Amsterdam, The Netherlands} \\
{\tt J.Smolic@uva.nl, M.Smolic@uva.nl}}
\newcommand{\auth}{{\large Jelena Smolic${}^1$ and Milena Smolic${}^1$}}

\newcommand{\be}{\begin{equation}} 
\newcommand{\ee}{\end{equation}} 
\newcommand{\bea}{\begin{eqnarray}} 
\newcommand{\eea}{\end{eqnarray}} 
\newcommand{\nn}{\nonumber}
\def\half{\frac{1}{2}}
\def\adot{\dot{\alpha}}
\def\bdot{\dot{\beta}}
\def\rdot{\dot{\rho}}

\def\gdot{\dot{\gamma}}
\def\a{\alpha}
\def\b{\beta}
\def\s{\sigma}
\def\S{\Sigma}
\def\l{\lambda}
\def\k{\kappa}
\def\r{\rho}
\def\d{\delta}
\def\t{\tau}

\def\g{\gamma}

\def\ve{\varepsilon}

\def\vp{\varphi}

\def\lbar{\bar{\lambda}}
\def\ebar{\bar{\eta}}
\def\ldot{\dot{\lambda}}
\def\sbar{\bar{\sigma}}

\def\stilde{\tilde{\sigma}}
\def\la{\langle}
\def\ra{\rangle}
\def\tri{\triangle}
\def\part{\partial}
\def\({\left(}
\def\){\right)}
\def\C{{\cal C}}

\def\mt{\tilde{M}}
\def\at{\tilde{A}}
\def\lt{\tilde{\lambda}}
\def\et{\tilde{\eta}}

\begin{document}




\vspace{15pt}

\begin{center}

{\Large \bf 2PI Effective Action and Evolution Equations of ${\cal N}$ = 4 super Yang-Mills}

\vspace{20pt}

\auth

\vspace{15pt}

\vspace{8pt}

{\ams}

\vspace{15pt}

\underline{ABSTRACT}
\end{center}
We employ $n$PI  effective action techniques to study ${\cal N}=4$ super Yang-Mills, and write down the 2PI effective action of the theory to two-loop order in the symmetric phase. We also supply the evolution equations of two-point correlators within the theory.

{\vfill\leftline{}\vfill
\pagebreak
\setcounter{page}{1}



\section{Introduction}
Our current understanding of phenomena in thermal equilibrium is extensive, but many of the processes which give deeper insights into our fundamental understanding of nature begin far away from equilibrium. There is abundant experimental data concerning the early stages of heavy-ion collisions, which requires the development of a nonequilibrium theoretical framework to allow for correct interpretation of the data. Furthermore, we might be able to make contact with the study of black hole formation and evaporation (see, for example \cite{Chesler:2008hg}) by using the AdS/CFT correspondence \cite{Maldacena:1997}. This widely studied correspondence postulates an exact equivalence between string theory on the $AdS_{5}\times S^{5}$ background and $(3+1)$-dimensional ${\cal N}=4$ super Yang-Mills theory (SYM) (for comprehensive reviews, see \cite{Aharony:1999} and \cite{Freedman:2002}). It is the best understood example of a gauge/gravity theory duality, i.e. of holography, a groundbreaking hypothesis which says that any gravitational theory should have a description in terms of a QFT with no gravity in one less dimension. It is precisely this ${\cal N}=4$ SYM which we wish to study further in a nonequilibrium setting. This is a very special theory from many perspectives, both because of the duality but also because intrinsically it has a lot of supersymmetry, and is considered the simplest QFT \cite{Nima:2010}. Thus, many things which are difficult to compute in conventional theories may be simpler to handle in this theory.

The problem of studying nonequilibrium phenomena is two-fold, because we not only need to take into account quantum fluctuations, but we also need to deal with a very large number of degrees of freedom. Classical statistical field theory simply is not good enough, and standard perturbative approaches based on small deviations from equilibrium are not applicable: secular, time-dependent terms may appear which invalidate the perturbative expansion. For example, in \cite{Hong:2006} it was argued that for $SU(N)$ Yang-Mills on a sphere, the high temperature phase of the theory is intrinsically non-perturbative. In recent years, so-called $n$PI effective action techniques have been developed \cite{Luttinger:1960, Baym:1962, Cornwall:1974, Smit:2002, Berges:2004_1}, allowing us to use nonperturbative approximations to get a handle on nonequilibrium dynamics, in the hope of ensuring non-secular and universal behaviour (meaning that the initial conditions do not affect the late-time behaviour). Of particular interest is the precise time evolution of quantum fields whose initial state is far from equilibrium. Relevant references relating to far-from-equilibrium quantum fields and thermalization include \cite{Calzetta:1988, Ivanov:1999, Berges:2000, Aarts:2001, Berges:2002, Ahrensmeier:2002, Serreau:2002, Juchem:2004, Arriz:2004, Tranberg:2005}. For a comprehensive review on progress in nonequilibrium QFT, see \cite{Berges:2004} and the references therein, and the book \cite{Hu:2008}.

Now, one can avoid having to calculate the full $n$PI effective action due to a useful equivalence hierarchy, which states that for a $q$-loop approximation all $n$PI descriptions with $n\geq q$ are equivalent, so only the $q$PI effective action is required. Thus, for example, a self-consistent description to two-loop order requires a 2PI effective action. For theories involving scalars and fermions, observing the approach to thermal equilibrium requires the three-loop 2PI effective action since the two-loop level involves an infinite number of spurious conserved quantities \cite{Berges:2004}. Furthermore, previous transport coefficient computations in QED and QCD have highlighted a number of issues with the 2PI approach in the context of gauge theories, which are resolved by going to 3PI level \cite{Berges:2004_1, Aarts:2003, Aarts:2005, 3loop1, 3loop7, 3loop8, 3loop10, 3loop11, 3loop12}. Nevertheless, we would like to present here the two-loop 2PI computation for ${\cal N}=4$ SYM, since this is already a highly nontrivial computation; our results may be viewed as a nontrivial stepping stone towards the full 3PI result. 

This paper is arranged as follows. In section 2 we present our results: we give the 2PI effective action of ${\cal N}=4$ SYM to two-loop order in the symmetric phase, and also write down the evolution equations of the 2-point correlators for the scalars, gluons, fermions and ghosts within the theory, in the nonequilibrium realm. In section 3 we discuss the issues which arise when the 2PI approach is applied to gauge theories. In section 4 we give the discussion and conclusions.

\section{The $n$PI Effective Action Approach}

In order to study nonequilibrium quantum field theory, one first needs to specify an initial state at some time $t_{0}$, which is usually done by specifying a density matrix $\r_{D}(t_{0})$ which is not a thermal equilibrium density matrix. Equivalently one can specify all initial $n$-point correlation functions, although in practice one often supplies only the lowest correlation functions at $t_{0}$. The time evolution of this initial state (i.e. these initial correlation functions) is then encoded in the functional path integral with classical action $S$. For example, in the case of a real scalar field $\varphi$, the nonequilibrium generating functional for correlation functions is
\be\label{eq:genfn}
Z[J_{1},J_{2},\cdots;\r_{D}] = Tr\left \{ \r_{D}(0) T_{{\cal C}} e^{i \( \int_{x} J_{1}(x)\Phi(x)+\half\int_{xy}J_{2}(x,y)\Phi(x)\Phi(y) +\ \cdots \)}\right \},
\ee
where $\Phi(x)$ denote Heisenberg field operators. $T_{{\cal C}}$ denotes time-ordering along the time path ${\cal C}$ and in what follows, $\int_{x}\equiv\int_{{\cal C}}dx^{0}\int d^{d}x$. It turns out that the extension to the nonequilibrium realm is done precisely via the introduction of this finite, closed real-time contour ${\cal C}$, known as the Schwinger-Keldysh contour, given in Figure \ref{fig:contour}.
\begin{figure}[hbt]
\centerline{ \epsfxsize 3in \epsfbox {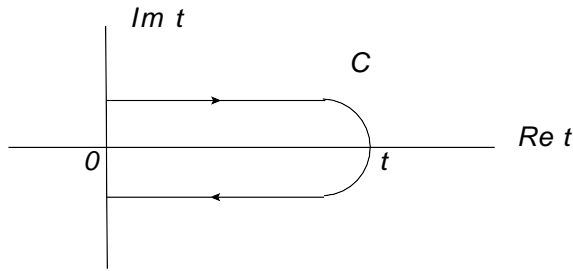} } \caption[]{
{\footnotesize Schwinger-Keldysh contour}.
} \label{fig:contour}
\end{figure}
We call the top part of the contour, the forward piece, ${\cal C}^{+}$, and the bottom backward piece ${\cal C}^{-}$. Time ordering along the contour is largely intuitive: we want any time on ${\cal C}^{-}$ to be later than any time on ${\cal C}^{+}$, so we use normal time ordering on the forward piece, and antitemporal time ordering along the backward piece. Time integration along the contour is given by
\be
\int_{{\cal C}} dx^{0} = \int_{0,{\cal C}^{+}}^{t} dx^{0} + \int_{t}^{0,{\cal C}^{-}} dx^{0} = \int_{0,{\cal C}^{+}}^{t}dx^{0} - \int_{0,{\cal C}^{-}}^{t}dx^{0}. 
\ee

In (\ref{eq:genfn}) we have included $n$ source terms $J_{1}(x)$, $J_{2}(x,y),\ \cdots$. Now, the standard prescription for writing down a 1PI (one particle irreducible) effective action involves introducing only the $J_{1}$ source term when setting up the generating functional, but in order to extend this to the so-called ``$n$PI" ($n$ particle irreducible) effective action, we need to introduce $n$ source terms. In practice, an equivalence hierarchy exists between $n$PI effective actions, and means that one can avoid having to calculate the effective action for arbitrarily large $n$. In our analysis of ${\cal N}=4$ SYM we will be interested in a two-loop approximation, which amounts to calculating the 2PI effective action of the theory. Studies of simpler theories, such as in \cite{Berges:2000} - \cite{Aarts:2005}, indicate that the three-loop 2PI effective action is necessary to see thermalization. For instance, in \cite{Aarts:2003, Aarts:2004, Aarts:2005} transport coefficients were computed correctly at leading order using the three-loop effective action. There are known difficulties in using the 2PI approach for gauge theories in the description of transport coefficients \cite{3loop1}, and much  progress in this direction has been made by extending the analysis to the 3PI effective action approach \cite{3loop7, 3loop8, 3loop10, 3loop11}. This is also likely necessary for ${\cal N}=4$ SYM, and these issues should be addressed, but as a first step we present the two-loop 2PI effective action. In (\ref{eq:genfn}) this would amount to including only the sources $J_{1}(x)$ and $J_{2}(x,y)$, so that we can rewrite the two-source generating functional as
\bea\label{eq:2piscalar}
Z[J_{1},J_{2};\r_{D}] &=&  \int d\vp^{+}d\vp^{-}\la \vp^{+}\vert \r_{D}(0)\vert\vp^{-} \ra\int_{\vp^{+}}^{\vp^{-}} {\cal D}^{\prime}\vp e^{i\(S[\vp]+\int_{x}J_{1}(x)\vp(x)+\half\int_{xy}J_{2}(x,y)\vp(x)\vp(y)\)}, \nn\\
&& \mbox{\ \ }
\eea
where $\vp^{\pm}$ are eigenstates of the Heisenberg field operators at initial time, namely $\Phi(t=0,\vec{x})\vert \vp^{\pm}\ra = \vp^{\pm}(\vec{x})\vert\vp^{\pm}\ra$. It follows from (\ref{eq:2piscalar}) that the structure of the nonequilibrium partition function is very similar to the zero temperature and thermal case, apart from the additional piece coming from the initial conditions, and the time ordering along ${\cal C}$ replacing the original time ordering along only ${\cal C}^{+}$. In principle, barring the initial conditions, we should be able to do our calculations at zero temperature, and easily transform our results to fit the nonequilibrium case by introducing the Schwinger-Keldysh contour. Indeed, notice that in (\ref{eq:2piscalar}) the second integral is basically just the vacuum generating functional for connected Green's functions for a scalar field theory with classical action $S[\varphi]$, in the presence of two source terms,
\bea
Z[J_{1},J_{2}]&=&e^{iW[J_{1},J_{2}]}=\int{\cal D}\vp\ exp\( i\left [ S[\vp] + \int_{x}J_{1}(x)\vp(x) +\frac{1}{2} \int_{xy}J_{2}(x,y)\vp(x)\vp(y)\right ] \). \nn\\
&&\mbox{\ \ }
\eea
 
Now, in order to obtain the equations of motion of the correlation functions of such a simple scalar field theory, thereby fully describing it, one first extracts the effective action $\Gamma$ from $S$ by performing suitable Legendre transforms. In addition to the Legendre transform of the generating functional necessary to obtain the 1PI effective action, we simultaneously perform a second Legendre transform to get the required 2PI effective action, namely
\bea\label{eq:2pieffactsc}
\Gamma[\phi,G] &=& W[J_{1},J_{2}]-\int_{x}\frac{\d W[J_{1},J_{2}]}{\d J_{1}(x)}J_{1}(x) - \int_{xy}\frac{\d W[J_{1},J_{2}]}{\d J_{2}(x,y)}J_{2}(x,y) \nn\\
&=& W[J_{1},J_{2}]-\int_{x}\phi(x)J_{1}(x)-\half\int_{xy}J_{2}(x,y)\phi(x)\phi(y)-\half Tr GJ_{2}.
\eea
Here $\phi(x)$ is the scalar field expectation value given by
\bea\label{eq:scfld}
\phi(x) =\frac{\d W[J_{1},J_{2}]}{\d J_{1}(x)},
\eea
and $G(x,y)$ is the connected two-point function. An $n$PI effective action would require $n$ simultaneous Legendre transforms of this type. Finally, first order functional derivatives of this effective action with respect to the appropriate correlation functions (in the absence of sources) then provide the corresponding equations of motion via the so-called ``stationarity conditions",
\be
\frac{\d\Gamma[\phi,G]}{\d\phi}=0,\ \ \ \frac{\d\Gamma[\phi,G]}{\d G}=0.
\ee

For more on the 2PI effective action of scalar and fermion fields, see \cite{Luttinger:1960, Baym:1962, Cornwall:1974, Berges:2004} and the references therein.

\subsection{Two-loop 2PI Effective Action of ${\cal N}=4$ SYM}

${\cal N}$ = 4 super Yang-Mills (SYM) has an $SU(N)$ colour gauge symmetry and corresponding gauge field $A_{\mu}$, and also contains four spinors $\l_{i}$, where $i = 1, \ldots , 4$ transforming under the global $SU(4)$ symmetry, and six scalars $M^{m}$, where $m = 1, \ldots , 6$, transforming under $SO(6)$. In order to quantize this theory properly we need to gauge-fix, which we do using the Faddeev-Popov procedure. Together with the gauge-fixing term $-\frac{1}{2\xi}Tr(\part_{\mu}A^{\mu})^{2}$ and ghost term  $Tr(-\ebar\part^{\mu}(\nabla_{\mu}\eta))$ (with ghost fields labelled by $\eta$ and anti ghosts by $\ebar$), the ${\cal N}=4$ SYM action is given by
\be
S_{SYM} = S_{SYM}^{0} + S_{SYM}^{int},
\ee
where
\be
S_{SYM}^{0} = \int_{x} Tr \( \half A_{\mu} \part^{2} A^{\mu} + \half M_{m}\part^{2}M^{m} + i\lbar^{\adot i} \sbar_{\adot\b}^{\mu}\part_{\mu} \l_{i}^{\b} -\ebar \part^{2}\eta \), 
\ee
and
\bea
S_{SYM}^{int} &=& \int_{x}Tr\( -ig\( \part_{\mu}A_{\nu}A^{\mu}A^{\nu}-\part_{\mu}A_{\nu}A^{\nu}A^{\mu}\right ) +\half g^{2}\left ( A_{\mu}A_{\nu}A^{\mu}A^{\nu}-A_{\mu}A_{\nu}A^{\nu}A^{\mu}\) \right .  \nn\\
&-&\left . ig\left ( \part_{\mu}M_{m}A^{\mu}M^{m}-\part_{\mu}M_{m}M^{m}A^{\mu}\right ) + g^{2}\left ( A_{\mu}M_{m}A^{\mu}M^{m}-M_{m}A_{\mu}A^{\mu}M^{m}\right )\right . \nn\\
&+&\left . \half g^{2} \left ( M_{m}M_{n}M^{m}M^{n}-M_{m}M_{n}M^{n}M^{m}\right)- g\left (\l_{i}^{\a}\s_{\a\bdot}^{\mu}A_{\mu}\lbar^{\bdot i}-\l_{i}^{\a}\s_{\a\bdot}^{\mu}\lbar^{\bdot i}A_{\mu}\right) \right . \nn\\
&+&\left .  \half i g \left ( \l_{i}^{\a}\l_{\a j}(\stilde_{m}^{-1})^{ij}M_{m} - \l_{i}^{\a}(\stilde_{m}^{-1})^{ij}M_{m}\l_{\a j} -\lbar_{\adot}^{i}\lbar^{\adot j}(\stilde_{m})_{ij}M_{m}+\lbar_{\adot}^{i}(\stilde_{m})_{ij}M_{m}\lbar^{\adot j} \right )\right . \nn\\
&+&\left .  ig \(\part^{\mu}\ebar A_{\mu} \eta - \part^{\mu}\ebar \eta A_{\mu}\) \). \nn\\
&&\mbox{\ \ }
\eea
In the above, $\nabla_{\mu}$ is the covariant derivative, with $\nabla_{\mu}\eta = \part_{\mu}\eta + ig[A_{\mu},\eta]$, and we work in the Feynman 'tHooft gauge where $\xi=1$. The $\it{free}$ gluon, scalar, fermion and ghost propagators (denoted $D_{0}, S_{0}, \tri_{0}$, and $ G_{0}$ respectively) are given by
\be
D_{0}(x,y) = \frac{i}{4\pi^{2}}\frac{1}{(x-y)^{2}} = S_{0}(x,y) = -G_{0}(x,y) ,\ \ \ \ \ \tri_{0,a\bdot}(x-y) = i\s_{\a\bdot}^{\mu}\part_{\mu}^{x}D_{0}(x,y).
\ee

The 2PI effective action of ${\cal N}=4$ SYM is given by 
\bea\label{eq:effact}
\Gamma[\mt,\at,\lt,\et,S,D,\tri,G] &=& S_{SYM}[\mt,\at,\lt,\et] \nn\\
&+& \frac{i}{2}TrlnS^{-1} + \frac{i}{2}TrS_{0}^{-1}S+ \frac{i}{2}TrlnD^{-1} + \frac{i}{2}TrD_{0}^{-1}D \nn\\
&-& iTrln\tri^{-1} -iTr\tri_{0}^{-1}\tri -iTrlnG^{-1} -iTr G_{0}^{-1}G \nn\\
&+& \Gamma_{2}[\mt,\at,\lt,\et,S,D,\tri,G], 
\eea
where the $S$, $D$, $\tri$ and $G$ are $\it{full}$ scalar, gluon, fermion and ghost propagators respectively, and $\mt$, $\at$, $\lt$, $\et$ are the respective field expectation values defined analogously to (\ref{eq:scfld}). The first nine terms above represent the 2PI effective action to one loop order, while $\Gamma_{2}[\mt,\at,\lt,\et,S,D,\tri,G]$ is the higher loop contribution. $\Gamma_{2}[\mt,\at,\lt,\et,S,D,\tri,G]$ is obtained by shifting each of the fields in $S_{SYM}$ by the respective field expectation value, and using the vertices obtained from this shifted action to build the higher loop 2PI diagrams. We will be working in the symmetric phase, $\it{i.e.}$ we set all field expectation values to zero. 

At two-loop order, the eight diagrams given in Figure \ref{fig:diagrams} contribute to $\Gamma_{2}$ in (\ref{eq:effact}).
\begin{figure}[hbt]
\centerline{ \epsfxsize 6in \epsfbox {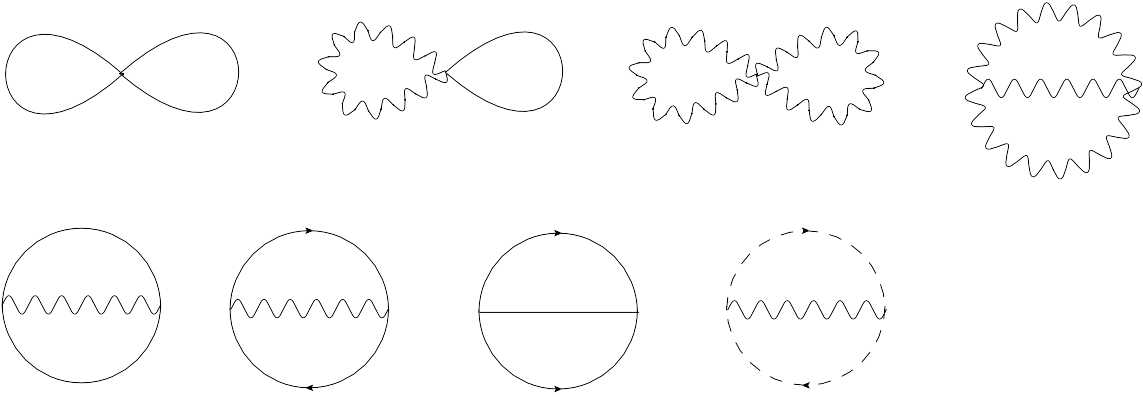} } \caption[0.5]{
{\footnotesize 2PI 2-loop diagrams contributing to $\Gamma_{2} [\mt,\at,\lt,\et,S,D,\tri,G]$ in the symmetric phase. Starting at the top row and going from left to right, we label them as $S^{2}$, $DS$, $D^{2}$, $D^{3}$, $S^{2}D$, $\tri\bar{\tri} D$, $\tri^{2}S$, and $G\bar{G}D$}.
} \label{fig:diagrams}
\end{figure}
The wavy lines correspond to gluons, straight lines to scalars, arrowed lines to fermions and dashed arrowed lines to ghosts. We will label the eight diagrams according to the propagators they contain. Thus, starting from the top row in Figure \ref{fig:diagrams} and going from left to right, we have $S^{2}$, $DS$, $D^{2}$, $D^{3}$, $S^{2}D$, $\tri\bar{\tri} D$, $\tri^{2}S$, and $G\bar{G}D$. Thus, 
\be
\Gamma_{2}[S,D,\tri,G]_{SYM} = \Gamma_{S^{2}} + \Gamma_{DS} + \Gamma_{D^{2}} + \Gamma_{D^{3}} + \Gamma_{S^{2}D} + \Gamma_{\tri\bar{\tri} D} + \Gamma_{\tri^{2}S} + \Gamma_{G\bar{G}D}, \nn
\ee
and
\bea
\Gamma_{S^{2}} &=& -15 g^{2}(N^{3}-N)\int_{x}S^{2}(x,x), \nn\\
\Gamma_{DS} &=& -6g^{2}(N^{3}-N)\int_{x}D_{\mu}^{\mu}(x,x)S(x,x), \nn\\
\Gamma_{D^{2}} &=& \half g^{2}(N^{3}-N)\int_{x}\(D_{\mu}^{\nu}(x,x)D_{\nu}^{\mu}(x,x) - D_{\mu}^{\mu}(x,x)D_{\nu}^{\nu}(x,x)\), \nn\\
\Gamma_{D^{3}} &=& -ig^{2}(N^{3}-N)\int_{xy}\( \part_{\mu}^{x}\part_{\r}^{y}D_{\nu\k}(x,y)\( D^{\mu\k}(x,y)D^{\nu\r}(x,y) - D^{\mu\r}(x,y)D^{\nu\k}(x,y)\) \right . \nn\\
&&\left . +\part_{\mu}^{x}D_{\nu}^{\r}(x,y)\(\part_{\r}^{y}D_{\k}^{\mu}(x,y)D^{\nu\k}(x,y) - \part_{\r}^{y}D_{\k}^{\nu}(x,y)D^{\mu\k}(x,y)\) \right . \nn\\
&&\left . +\part_{\mu}^{x}D_{\nu}^{\k}(x,y)\(\part_{\r}^{y}D_{\k}^{\nu}(x,y)D^{\mu\r}(x,y) - \part_{\r}^{y}D_{\k}^{\mu}(x,y)D^{\nu\r}(x,y) \) \), \nn\\
\Gamma_{S^{2}D} &=& -6ig^{2}(N^{3}-N)\int_{xy}D^{\mu\r}(x,y)\( \part_{\mu}^{x}S(x,y)\part_{\r}^{y}S(x,y)-\part_{\mu}^{x}\part_{\r}^{y}S(x,y)S(x,y) \), \nn\\
\Gamma_{\tri\bar{\tri}D} &=& -4ig^{2}(N^{3}-N)\s_{\a\bdot}^{\mu}\s_{\k\rdot}^{\nu}\int_{xy}\tri^{\a\rdot}(x,y)\tri^{\k\bdot}(y,x)D_{\mu\nu}(x,y), \nn
\eea
\bea
&&\mbox{\ } \nn\\
\Gamma_{\tri^{2}S} &=& -24ig^{2}(N^{3}-N)\ve_{\a\b}\ve_{\bdot\adot}\int_{xy}\tri^{\a\adot}(x,y)\tri^{\b\bdot}(x,y)S(x,y), \nn\\
\Gamma_{G\bar{G}D} &=& ig^{2} (N^{3}-N)\int_{xy}\part_{x}^{\mu}G(y,x)\part_{y}^{\nu}G(x,y)D_{\mu\nu}(x,y),
\eea
where we emphasize again that all the propagators above are $\it{full}$. As a check, we can evaluate the two-loop 2PI effective action of ${\cal N}=4$ SYM obtained above to ${\cal O}(g^{2})$, by substituting the free propagators when evaluating each of the diagrams. Due to the conformal and supersymmetric nature of ${\cal N}=4$ SYM, we expect the effective action to vanish, and it does. This is a first check that our effective action is indeed correct. 

\subsection{Evolution Equations of ${\cal N}=4$ SYM}

Having determined which diagrams contribute to the two-loop 2PI effective action of ${\cal N}=4$ SYM, we can use the stationarity conditions to write down the equations of motion for each of the fields in our theory \cite{Berges:2002, Berges:2004}. This involves writing down the self energy $\S(x,y)$ for each propagator in the usual way (by cutting that propagator line on each of the 2PI diagrams at our disposal). As mentioned in the beginning of section 2, barring the initial conditions, we should be able to do our calculations in the vacuum, and easily transform our results to fit the nonequilibrium case by introducing the Schwinger-Keldysh contour.

For scalars, the equations of motion are
\bea\label{eq:scalar equations}
\(\part_{x}^{2} -\S^{(0)(s)}(x;S)\)F^{(s)}(x,y)&=&\int_{0}^{x^{0}}dz^{0}\int d^{3}z \S_{\r}^{(s)}(x,z)F^{(s)}(z,y) \nn\\
&-& \int_{0}^{y^{0}}dz^{0}\int d^{3}z \S_{F}^{(s)}(x,z)\r^{(s)}(z,y), \nn\\
\(\part_{x}^{2} -\S^{(0)(s)}(x;S)\)\r^{(s)}(x,y)&=&\int_{y^{0}}^{x^{0}}dz^{0}\int d^{3}z \S_{\r}^{(s)}(x,z)\r^{(s)}(z,y). 
\eea
The superscript $(s)$ refers to scalars (and similarly, throughout this paper, the superscripts $(gl)$=gluon, $(f)$=fermion and $(gh)$=ghost).

The integro-differential evolution equations in (\ref{eq:scalar equations}) are for the scalar statistical propagator $F^{(s)}(x,y)$ and the scalar spectral function $\r^{(s)}(x,y)$, which are obtained via a splitting of the scalar two-point function \cite{Aarts:2001, Berges:2002}
\be
S_{{\cal C}}(x,y) = F^{(s)}(x,y) - \frac{i}{2}\r^{(s)}(x,y)\mbox{sign}_{{\cal C}}(x^{0}-y^{0}).
\ee
The preference in using $F(x,y)$ and $\r(x,y)$ is that they are both real, which makes their evolution equations intrinsically more manageable, and more importantly they have handy physical interpretations. The spectral function involves the spectrum of the theory while the statistical propagator deals with occupation numbers.

The quantities $\S_{F}^{(s)}(x,y)$ and $\S_{\r}^{(s)}(x,y)$ arise from a similar splitting of the scalar self-energy, namely
\be
\S_{\C}^{(s)}(x,y) = \d_{{\cal C}}(x,y)\S^{(0)(s)}(x) + \S_{F}^{(s)}(x,y) - \frac{i}{2}\S_{\r}^{(s)}(x,y)\mbox{sign}_{{\cal C}}(x^{0}-y^{0}).
\ee
The local term $\S^{(0)(s)}(x)$ is part of a generalized ``mass" term. (When considering the evolution equations for the other fields in ${\cal N}=4$ SYM, we perform a similar splitting of the two-point functions and corresponding self-energies, and label them by the superscripts $(gl)$, $(f)$ and $(gh)$).

Of course, the scalar evolution equations in (\ref{eq:scalar equations}) are quite general. In order for them to apply specifically to ${\cal N}=4$ SYM, we need to write down the quantities $\S_{F}^{(s)}$ and $\S_{\r}^{(s)}$ in the context of this theory. These quantities are related to the corrections to the scalar propagator, which we obtain by cutting a scalar line in the two-loop diagrams of Figure \ref{fig:diagrams}. The diagrams in Figure \ref{fig:scalar corr} contribute. 
\begin{figure}[hbt]
\centerline{ \epsfxsize 6in \epsfbox {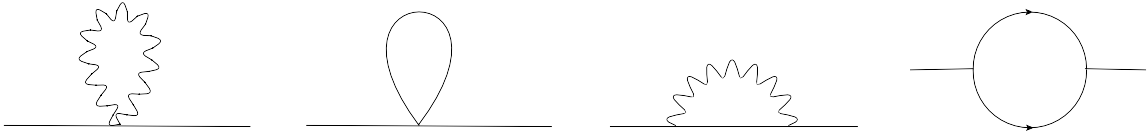} } \caption[]{
{\footnotesize Corrections to the scalar propagator.} 
} \label{fig:scalar corr}
\end{figure}
Being careful to take the contour $\C$ into account and suppressing adjoint indices, we obtain

\bea\label{eq:sigfscalar}
\S_{F}^{(s)}(x,z)&=& -g^{2}N \left [ 8 \part_{\nu}^{z}\part_{\mu}^{x}F^{(s)}(x,z)F^{(gl)\mu\nu}(x,z) -2 \part_{\nu}^{z}\part_{\mu}^{x}\r^{(s)}(x,z)\r^{(gl)\mu\nu}(x,z) \right . \nn\\
&&\qquad\ \ + \left . 4 \part_{\mu}^{x}F^{(s)}(x,z)\part_{\nu}^{z}F^{(gl)\mu\nu}(x,z) - \part_{\mu}^{x}\r^{(s)}(x,z)\part_{\nu}^{z}\r^{(gl)\mu\nu}(x,z) \right . \nn\\
&&\qquad\ \ + \left . 4\part_{\nu}^{z}F^{(s)}(x,z)\part_{\mu}^{x}F^{(gl)\mu\nu}(x,z) - \part_{\nu}^{z}\r^{(s)}(x,z)\part_{\mu}^{x}\r^{(gl)\mu\nu}(x,z) \right . \nn\\
&&\qquad\ \ + \left .  2 F^{(s)}(x,z)\part_{\nu}^{z}\part_{\mu}^{x}F^{(gl)\mu\nu}(x,z) - \half \r^{(s)}(x,z)\part_{\nu}^{z}\part_{\mu}^{x}\r^{(gl)\mu\nu}(x,z) \right . \nn\\
&&\qquad\ \ -\left .  \ve_{\a\b}\ve_{\bdot\adot} \( 4 F^{(f)\a\adot}(x,z)F^{(f)\b\bdot}(x,z) - \r^{(f)\a\adot}(x,z)\r^{(f)\b\bdot}(x,z) \right. \right . \nn\\
&&\qquad\ \ + \left . \left . 4F^{(f)\a\adot}(z,x)F^{(f)\b\bdot}(z,x) - \r^{(f)\a\adot}(z,x)\r^{(f)\b\bdot}(z,x) \) \right ], \nn\\
&&\mbox{\ \ }
\eea
\bea\label{sigrscalar}
\S_{\r}^{(s)}(x,z) &=& -g^{2}N \left [ 8 \( \part_{\nu}^{z}\part_{\mu}^{x}F^{(s)}(x,z)\r^{(gl)\mu\nu}(x,z) -2 \part_{\nu}^{z}\part_{\mu}^{x}\r^{(s)}(x,z)F^{(gl)\mu\nu}(x,z)\) \right . \nn\\
&&\qquad\ \ + \left . 4 \(\part_{\mu}^{x}F^{(s)}(x,z)\part_{\nu}^{z}\r^{(gl)\mu\nu}(x,z) - \part_{\mu}^{x}\r^{(s)}(x,z)\part_{\nu}^{z}F^{(gl)\mu\nu}(x,z)\) \right . \nn\\
&&\qquad\ \ + \left . \(4\part_{\nu}^{z}F^{(s)}(x,z)\part_{\mu}^{x}\r^{(gl)\mu\nu}(x,z) - \part_{\nu}^{z}\r^{(s)}(x,z)\part_{\mu}^{x}F^{(gl)\mu\nu}(x,z)\) \right . \nn\\
&&\qquad\ \ + \left .  2 F^{(s)}(x,z)\part_{\nu}^{z}\part_{\mu}^{x}\r^{(gl)\mu\nu}(x,z) + 2 \r^{(s)}(x,z)\part_{\nu}^{z}\part_{\mu}^{x}F^{(gl)\mu\nu}(x,z) \right . \nn\\
&&\qquad\ \ -\left .  4 \ve_{\a\b}\ve_{\bdot\adot} \( F^{(f)\a\adot}(x,z)\r^{(f)\b\bdot}(x,z) + \r^{(f)\a\adot}(x,z)F^{(f)\b\bdot}(x,z) \right. \right . \nn\\
&&\qquad\ \ - \left . \left . F^{(f)\a\adot}(z,x)\r^{(f)\b\bdot}(z,x) - \r^{(f)\a\adot}(z,x)F^{(f)\b\bdot}(z,x) \) \right ]. \nn\\
&&\mbox{\ \ }
\eea

With analogous definitions for the gluons, fermions and ghosts, we can also write down the evolution equations for the statistical propagators and spectral functions of these fields.
For gluons, the equations are
\bea
\(g_{\nu}^{\k}\part_{x}^{2} -\S_{\ \ \ \ \ \ \nu}^{(0)(gl)\k}(x;D)\)F^{(gl)\nu\g}(x,y)&=&\int_{0}^{x^{0}}dz^{0}\int d^{3}z \S_{\r,\ \nu}^{(gl)\k}(x,z)F^{(gl)\nu\g}(z,y) \nn\\
&-& \int_{0}^{y^{0}}dz^{0}\int d^{3}z \S_{F,\ \nu}^{(gl)\k}(x,z)\r^{(gl)\nu\g}(z,y), \nn\\
\(g_{\nu}^{\k}\part_{x}^{2} -\S_{\ \ \ \ \ \ \nu}^{(0)(gl)\k}(x;D)\)\r^{(gl)\nu\g}(x,y)&=&\int_{y^{0}}^{x^{0}}dz^{0}\int d^{3}z \S_{\r,\ \nu}^{(gl)\k}(x,z)\r^{(gl)\nu\g}(z,y). \nn\\
&&\mbox{\ \ } 
\eea
The fermion equations are
\bea
\( i\sbar_{\adot\b}^{\mu}\part_{\mu}^{x} + \S_{\adot\b}^{(0)(f)}(x;\tri) \)F^{(f)\b\gdot}(x,y)&=&\int_{0}^{x^{0}}dz^{0}\int d^{3}z \S_{\r,\ \adot\b}^{(f)}(x,z)F^{(f)\b\gdot}(z,y) \nn\\
&& - \int_{0}^{y^{0}}dz^{0}\int d^{3}z \S_{F,\ \adot\b}^{(f)}(x,z)\r^{(f)\b\gdot}(z,y), \nn\\
\( i\sbar_{\adot\b}^{\mu}\part_{\mu}^{x} + \S_{\adot\b}^{(0)(f)}(x;\tri) \)\r^{(f)\b\gdot}(x,y)&=& \int_{y^{0}}^{x^{0}}dz^{0}\int d^{3}z\S_{\r,\ \adot\b}^{(f)}(x,z)\r^{(f)\b\gdot}(z,y).\nn\\
&&\mbox{\ \ }  
\eea
Finally, the ghost equations are
\bea
\(\part_{x}^{2} -\S^{(0)(gh)}(x;G)\)F^{(gh)}(x,y)&=&\int_{0}^{y^{0}}dz^{0}\int d^{3}z \S_{F}^{(gh)}(x,z)\r^{(gh)}(z,y) \nn\\
&-& \int_{0}^{x^{0}}dz^{0}\int d^{3}z \S_{\r}^{(gh)}(x,z)F^{(gh)}(z,y), \nn\\
\(\part_{x}^{2} -\S^{(0)(gh)}(x;G)\)\r^{(gh)}(x,y)&=&- \int_{y^{0}}^{x^{0}}dz^{0}\int d^{3}z \S_{\r}^{(gh)}(x,z)\r^{(gh)}(z,y). \nn\\
&&\mbox{\ \ } 
\eea
The corresponding $\S_{F}^{(gl)/(f)/(gh)}$ and $\S_{\r}^{(gl)/(f)/(gh)}$ for each of the equations above are given in the Appendix. 

\section{Issues related to the 2PI approach for gauge theories}

As mentioned before, due to a useful equivalence hierarchy a self consistent description to $q$-loop order requires a $q$PI effective action. Thus, working to two-loop order requires a 2PI effective action. A number of issues arise when the 2PI approach is applied to gauge theories. In particular, when using the 2PI approach to calculate transport coefficients in theories such as QED and QCD, it becomes clear that the importance of a diagram does not necessarily correlate with its loop order (we refer to \cite{Berges:2004_1, Aarts:2003, Aarts:2005, 3loop1, 3loop7, 3loop8, 3loop10, 3loop11, 3loop12} for full discussions). Soft and collinear momenta within gauge theories lead to enhancements which eliminate regular loop counting. In fact, one would need to include an infinite series of 2PI ``ladder" diagrams to recover the leading order ``on-shell" results for transport coefficients within QED or QCD (a manifestation of the LPM effect \cite{LPM1, LPM2, LPM3, LPM4, LPM5}). In addition to such collinear contributions, one also has to deal with ``pinch singularities" when calculating transport coefficients. These again produce an infinite set of diagrams which also need to be resummed. It turns out that the coupled integral equations which allow the resummation of the collinear and pinching terms and yield the complete leading order result for the transport coefficients (as obtained via kinetic theory), can be derived directly from the three-loop 3PI effective action. Since transport coefficients characterize equilibration in systems which are locally close to equilibrium and homogeneous on rather large scales, being able to compute them is the first step towards more advanced nonequilibrium calculations.

Thus, it appears that if we wish to make even the most basic transport coefficient computations for ${\cal N}=4$ SYM we need to begin with the 3PI effective action to three-loops. In fact, even in simpler scalar field theories involving scalars and fermions a two-loop 2PI approximation (Hartree, or similarly leading order large $N$ approximation) will not allow us to see thermalization, due to the presence of an infinite number of spurious conserved quantities \cite{Berges:2004}, and in order to study properly nonequilibrium evolution we would need to include the three-loop contribution. As a first step towards the full three-loop (3PI) result (which would be applicable to physical computations like that of transport coefficients), we begin by considering the two-loop computation. At this level, we require the 2PI effective action for self-consistency. Even to this order the evolution equations we obtain (Appendix) are fairly complicated, and even though our current result may not yet allow for nontrivial physical predictions, we consider our calculation the first nontrivial and necessary step on the way to a three-loop computation. 
 
\section{Discussion and conclusions}

In this paper we used the $n$PI effective action approach to write down the two-loop 2PI effective action of ${\cal N}=4$ SYM in the symmetric phase. We then wrote down the evolution equations for the two-point correlators of scalars, gluons, fermions and ghosts in the theory. A particularly pleasing property of ${\cal N}=4$ SYM is that it is a finite theory, so no renormalization is required. $n$PI effective actions enable us to set up a very efficient nonperturbative approximation scheme for nonequilibrium QFT, in the hope that we may bypass the usual problems of secularity and non-universality experienced by the standard perturbative approaches.

An important step in understanding the nonequilibrium dynamics of quantum fields is to understand how systems which are initially far from equilibrium approach thermal equilibrium at late times. We want to understand thermalization in QFTÕs which have a holographic dual, in particular ${\cal N}=4$ SYM. Much work has been done in studying transport coefficients using the $n$PI approach with the result that for gauge theories such as QED and QCD, one needs to use the three-loop 3PI formalism in order to reproduce the leading order results obtained from kinetic theory \cite{3loop7, 3loop8, 3loop10, 3loop11}. Thus, a first step in extending our results would be to push our calculation to three-loop order, which requires a 3PI effective action for self-consistency. The next step would then be to consider various initial conditions and solve the equations of motion, thereby allowing one to explore thermalization in this theory. Arguably the simplest possibility is to consider gaussian initial conditions and solve the equations of motion numerically (any such solution would have to be numerical, due to the extremely nonlinear and coupled nature of the evolution equations). One could then move on to more complicated initial conditions, since physical initial conditions may not be gaussian. In particular, since ${\cal N}=4$ SYM is a supersymmetric conformal field theory, supersymmetric initial conditions may simplify things. Since the 2PI effective action in general has a gauge dependence \cite{Smit:2002, 3loop4, 3loop9}, it would be interesting to see the dependence of our results on gauge fixing, especially when the calculation has been pushed to 3PI three-loop order. Resummed 2PI and 3PI effective actions in gauge theories have been used in \cite{3loop7, 3loop8, 3loop10, 3loop11, 3loop12} when calculating transport coefficients to ensure explicitly gauge invariant results, and further investigation into such a scheme for ${\cal N}=4$ SYM would aid in a generalization to nonabelian gauge theories. Transport coefficients of ${\cal N}=4$ SYM have been computed in \cite{Moore:2007} using kinetic theory and a comparison of this result with the result from the 3PI approach would provide a first check of the method. Exploring thermalization in this theory and comparing it to that in QCD would potentially give insights as to why the RHIC data seems to be well-described by the strongly coupled ${\cal N}=4$ theory. It would also be interesting to investigate the implications for black hole formation and evaporation, since, through holography, the process of thermalization is expected to be mapped to horizon formation on the gravitational side. In the context of holography, a similar formulation was developed in \cite{Balt:2008_1, Balt:2008_2, Balt:2009}, but  instead for gauge invariant operators without the $n$PI technique.

\section*{Acknowledgements}

We would like to thank K. Skenderis and M. Taylor for initially suggesting this problem and for many useful discussions. We would also like to thank J. Smit for discussions. This work is part of a research program which is financially supported by the `Nederlandse Organisatie voor Wetenschappelijk Onderzoek' (NWO). JS and MS acknowledge support via the NWO Vici grant of K. Skenderis.

\appendix
\section {Appendix}

The corrections to the gluon propagator in ${\cal N}=4$ SYM are given by the six diagrams in Figure \ref{fig:gluon corr}.
\begin{figure}[hbt]
\centerline{ \epsfxsize 6in \epsfbox {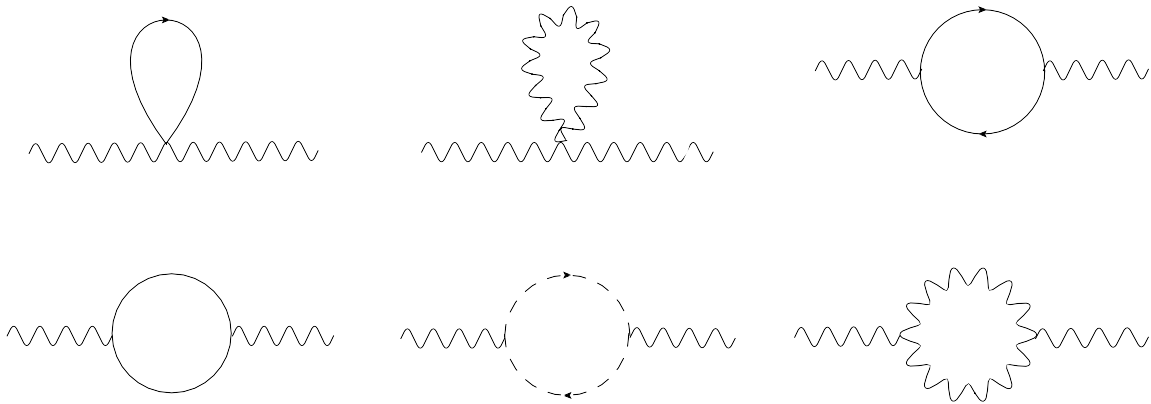} } \caption[]{
{\footnotesize Corrections to gluon propagator.}
} \label{fig:gluon corr}
\end{figure}
Thus, we have

\bea
\S_{F}^{(gl)\r\xi}(x,z) &=& g^{2}N \left [ \s_{\a\bdot}^{\r}\s_{\k\rdot}^{\xi}\(8F^{(f)\a\rdot}(x,z)F^{(f)\k\bdot}(z,x) +2\r^{(f)\a\rdot}(x,z)\r^{(f)\k\bdot}(z,x) \) \right . \nn\\
&& \left . \qquad+12 \part_{x}^{\r}F^{(s)}(x,z)\part_{z}^{\xi}F^{(s)}(x,z) -3 \part_{x}^{\r}\r^{(s)}(x,z)\part_{z}^{\xi}\r^{(s)}(x,z) \right . \nn\\
&& \left . \qquad-12 \part_{x}^{\r}\part_{z}^{\xi}F^{(s)}(x,z)F^{(s)}(x,z) +3 \part_{x}^{\r}\part_{z}^{\xi}\r^{(s)}(x,z)\r^{(s)}(x,z) \right . \nn\\
&& \left. \qquad-2 \part_{z}^{\xi}F^{(gh)}(x,z)\part_{x}^{\r}F^{(gh)}(z,x) -\half\part_{z}^{\xi}\r^{(gh)}(x,z)\part_{x}^{\r}\r^{(gh)}(z,x) \right . \nn\\
&& \left . \qquad-8 \( 2\part_{\mu}^{x}\part_{\b}^{z}F^{(gl)\r\xi}(x,z)F^{(gl)\mu\b}(x,z) - \half \part_{\mu}^{x}\part_{\b}^{z}\r^{(gl)\r\xi}(x,z)\r^{(gl)\mu\b}(x,z) \) \right . \nn\\
&& \left . \qquad+4 \( 2\part_{\mu}^{x}\part_{\b}^{z}F^{(gl)\r\b}(x,z)F^{(gl)\mu\xi}(x,z) - \half \part_{\mu}^{x}\part_{\b}^{z}\r^{(gl)\r\b}(x,z)\r^{(gl)\mu\xi}(x,z) \) \right . \nn\\
&& \left . \qquad+4 \( 2\part_{\mu}^{x}\part_{\b}^{z}F^{(gl)\mu\xi}(x,z)F^{(gl)\r\b}(x,z) - \half \part_{\mu}^{x}\part_{\b}^{z}\r^{(gl)\mu\xi}(x,z)\r^{(gl)\r\b}(x,z) \) \right . \nn\\
&& \left . \qquad-2 \( 2\part_{\mu}^{x}\part_{\b}^{z}F^{(gl)\mu\b}(x,z)F^{(gl)\r\xi}(x,z) - \half \part_{\mu}^{x}\part_{\b}^{z}\r^{(gl)\mu\b}(x,z)\r^{(gl)\r\xi}(x,z) \) \right . \nn\\
&& \left . \qquad+4 \( 2\part_{x}^{\mu}\part_{z}^{\xi}F^{(gl)\r\k}(x,z)F_{\mu\k}^{(gl)}(x,z) -\half \part_{x}^{\mu}\part_{z}^{\xi}\r^{(gl)\r\k}(x,z)\r_{\mu\k}^{(gl)}(x,z) \) \right . \nn\\
&& \left . \qquad+4 \( 2\part_{x}^{\r}\part_{z}^{\b}F^{(gl)\nu\xi}(x,z)F_{\nu\b}^{(gl)}(x,z) -\half \part_{x}^{\r}\part_{z}^{\b}\r^{(gl)\nu\xi}(x,z)\r_{\nu\r}^{(gl)}(x,z) \) \right . \nn\\
&& \left . \qquad-2 \( 2\part_{x}^{\r}\part_{z}^{\b}F_{\nu\b}^{(gl)}(x,z)F^{(gl)\nu\xi}(x,z) -\half \part_{x}^{\r}\part_{z}^{\b}\r_{\nu\b}^{(gl)}(x,z)\r^{(gl)\nu\xi}(x,z) \) \right . \nn\\
&& \left . \qquad-2 \( 2\part_{x}^{\r}\part_{z}^{\xi}F_{\nu\k}^{(gl)}(x,z)F^{(gl)\nu\k}(x,z) -\half \part_{x}^{\r}\part_{z}^{\xi}\r_{\nu\k}^{(gl)}(x,z)\r^{(gl)\nu\k}(x,z) \) \right . \nn\\
&& \left . \qquad-2 \( 2\part_{x}^{\mu}\part_{z}^{\xi}F_{\mu\k}^{(gl)}(x,z)F^{(gl)\r\k}(x,z) -\half \part_{x}^{\mu}\part_{z}^{\xi}\r_{\mu\k}^{(gl)}(x,z)\r^{(gl)\r\k}(x,z) \) \right . \nn\\
&& \left . \qquad+8 \(2\part_{\mu}^{x}F^{(gl)\r\b}(x,z)\part_{\r}^{z}F^{(gl)\mu\xi}(x,z) -\half\part_{\mu}^{x}\r^{(gl)\r\b}(x,z)\part_{\r}^{z}\r^{(gl)\mu\xi}(x,z) \) \right . \nn\\
&& \left . \qquad-4 \( 2\part_{x}^{\r}F_{\nu\b}^{(gl)}(x,z)\part_{z}^{\b}F^{(gl)\nu\xi}(x,z) - \half \part_{x}^{\r}\r_{\nu\b}^{(gl)}(x,z)\part_{z}^{\b}\r^{(gl)\nu\xi}(x,z) \) \right . \nn\\
&& \left . \qquad-4 \( 2\part_{x}^{\mu}F^{(gl)\r\k}(x,z)\part_{z}^{\xi}F_{\mu\k}^{(gl)}(x,z) -\half \part_{x}^{\mu}\r^{(gl)\r\k}(x,z)\part_{z}^{\xi}\r_{\mu\k}^{(gl)}(x,z) \) \right . \nn\\
&& \left . \qquad+2 \(2 \part_{x}^{\r}F^{(gl)\nu\k}(x,z)\part_{z}^{\xi}F_{\nu\k}^{(gl)}(x,z) -\half\part_{x}^{\r}\r^{(gl)\nu\k}(x,z)\part_{z}^{\xi}\r_{\nu\k}^{(gl)}(x,z) \) \right .  \nn\\
&& \left . \qquad-4 \( 2\part_{\mu}^{x}F^{(gl)\r\xi}(x,z)\part_{\b}^{z}F^{(gl)\mu\b}(x,z) -\half\part_{\mu}^{x}\r^{(gl)\r\xi}(x,z)\part_{\b}^{z}\r^{(gl)\mu\b}(x,z) \) \right . \nn\\
&& \left . \qquad+2 \( 2\part_{x}^{\r}F^{(gl)\nu\xi}(x,z)\part_{z}^{\b}F_{\nu\b}^{(gl)}(x,z) - \half\part_{x}^{\r}\r^{(gl)\nu\xi}(x,z)\part_{z}^{\b}\r_{\nu\b}^{(gl)}(x,z) \) \right . \nn\\
&& \left . \qquad-4 \( 2 \part_{\mu}^{x}F^{(gl)\mu\b}(x,z)\part_{\b}^{z}F^{(gl)\r\xi}(x,z) -\half \part_{\mu}^{x}\r^{(gl)\mu\b}(x,z)\part_{\b}^{z}\r^{(gl)\r\xi}(x,z) \) \right. \nn\\
&& \left . \qquad+2 \( 2\part_{x}^{\mu}F_{\mu\k}^{(gl)}(x,z)\part_{z}^{\xi}F^{(gl)\r\k}(x,z) -\half\part_{x}^{\mu}\r_{\mu\k}^{(gl)}(x,z)\part_{z}^{\xi}\r^{(gl)\r\k}(x,z) \) \right . \nn\\
&& \left . \qquad+2 \( 2\part_{\mu}^{x}F^{(gl)\mu\xi}(x,z)\part_{\b}^{z}F^{(gl)\r\b}(x,z) -\half \part_{\mu}^{x}\r^{(gl)\mu\xi}(x,z)\part_{\b}^{z}\r^{(gl)\r\b}(x,z) \) \right ], \nn\\
&&\mbox{\ \ } 
\eea

\bea
\S_{\r}^{(gl)\r\xi}(x,z) &=& g^{2}N\left [ \s_{\a\bdot}^{\r}\s_{\k\rdot}^{\xi} \( -8F^{(f)\a\rdot}(x,z)\r^{(f)\k\bdot}(z,x) + 8\r^{(f)\a\rdot}(x,z)F^{(f)\k\bdot}(z,x)\) \right . \nn\\
&& \left . \qquad+12\part_{x}^{\r}F^{(s)}(x,z)\part_{z}^{\xi}\r^{(s)}(x,z)+12\part_{x}^{\r}\r^{(s)}(x,z)\part_{z}^{\xi}F^{(s)}(x,z) \right . \nn\\
&& \left . \qquad-12\part_{x}^{\r}\part_{z}^{\xi}F^{(s)}(x,z)\r^{(s)}(x,z) -12\part_{x}^{\r}\part_{z}^{\xi}\r^{(s)}(x,z)F^{(s)}(x,z) \right . \nn\\
&& \left . \qquad+2\part_{z}^{\xi}F^{(gh)}(x,z)\part_{x}^{\r}\r^{(gh)}(z,x) - 2\part_{z}^{\xi}\r^{(gh)}(x,z)\part_{x}^{\r}F^{(gh)}(z,x) \right . \nn\\
&& \left . \qquad-16 \( \part_{\mu}^{x}\part_{\b}^{z}F^{(gl)\r\xi}(x,z)\r^{(gl)\mu\b}(x,z) + \part_{\mu}^{x}\part_{\b}^{z}\r^{(gl)\r\xi}(x,z)F^{(gl)\mu\b}(x,z) \) \right . \nn\\
&& \left . \qquad+8 \( \part_{\mu}^{x}\part_{\b}^{z}F^{(gl)\r\b}(x,z)\r^{(gl)\mu\xi}(x,z) +\part_{\mu}^{x}\part_{\b}^{z}\r^{(gl)\r\b}(x,z)F^{(gl)\mu\xi}(x,z) \) \right . \nn\\
&& \left . \qquad+8 \( \part_{\mu}^{x}\part_{\b}^{z}F^{(gl)\mu\xi}(x,z)\r^{(gl)\r\b}(x,z) + \part_{\mu}^{x}\part_{\b}^{z}\r^{(gl)\mu\xi}(x,z)F^{(gl)\r\b}(x,z) \) \right . \nn\\
&& \left . \qquad-4 \( \part_{\mu}^{x}\part_{\b}^{z}F^{(gl)\mu\b}(x,z)\r^{(gl)\r\xi}(x,z) +\part_{\mu}^{x}\part_{\b}^{z}\r^{(gl)\mu\b}(x,z)F^{(gl)\r\xi}(x,z) \) \right . \nn\\
&& \left . \qquad+8 \( \part_{x}^{\mu}\part_{z}^{\xi}F^{(gl)\r\k}(x,z)\r_{\mu\k}^{(gl)}(x,z)+ \part_{x}^{\mu}\part_{z}^{\xi}\r^{(gl)\r\k}(x,z)F_{\mu\k}^{(gl)}(x,z) \) \right . \nn\\
&& \left . \qquad+8 \( \part_{x}^{\r}\part_{z}^{\b}F^{(gl)\nu\xi}(x,z)\r_{\nu\b}^{(gl)}(x,z) + \part_{x}^{\r}\part_{z}^{\b}\r^{(gl)\nu\xi}(x,z)F_{\nu\r}^{(gl)}(x,z) \) \right . \nn\\
&& \left . \qquad-4 \( \part_{x}^{\r}\part_{z}^{\b}F_{\nu\b}^{(gl)}(x,z)\r^{(gl)\nu\xi}(x,z) + \part_{x}^{\r}\part_{z}^{\b}\r_{\nu\b}^{(gl)}(x,z)F^{(gl)\nu\xi}(x,z) \) \right . \nn\\
&& \left . \qquad-4 \( \part_{x}^{\r}\part_{z}^{\xi}F_{\nu\k}^{(gl)}(x,z)\r^{(gl)\nu\k}(x,z) + \part_{x}^{\r}\part_{z}^{\xi}\r_{\nu\k}^{(gl)}(x,z)F^{(gl)\nu\k}(x,z) \) \right . \nn\\
&& \left . \qquad-4 \( \part_{x}^{\mu}\part_{z}^{\xi}F_{\mu\k}^{(gl)}(x,z)\r^{(gl)\r\k}(x,z) +\part_{x}^{\mu}\part_{z}^{\xi}\r_{\mu\k}^{(gl)}(x,z)F^{(gl)\r\k}(x,z) \) \right . \nn\\
&& \left . \qquad+16 \(\part_{\mu}^{x}F^{(gl)\r\b}(x,z)\part_{\r}^{z}\r^{(gl)\mu\xi}(x,z) +\part_{\mu}^{x}\r^{(gl)\r\b}(x,z)\part_{\r}^{z}F^{(gl)\mu\xi}(x,z) \) \right . \nn\\
&& \left . \qquad-8 \( \part_{x}^{\r}F_{\nu\b}^{(gl)}(x,z)\part_{z}^{\b}\r^{(gl)\nu\xi}(x,z) +\part_{x}^{\r}\r_{\nu\b}^{(gl)}(x,z)\part_{z}^{\b}F^{(gl)\nu\xi}(x,z) \) \right . \nn\\
&& \left . \qquad-8 \( \part_{x}^{\mu}F^{(gl)\r\k}(x,z)\part_{z}^{\xi}\r_{\mu\k}^{(gl)}(x,z) + \part_{x}^{\mu}\r^{(gl)\r\k}(x,z)\part_{z}^{\xi}F_{\mu\k}^{(gl)}(x,z) \) \right . \nn\\
&& \left . \qquad+4 \( \part_{x}^{\r}F^{(gl)\nu\k}(x,z)\part_{z}^{\xi}\r_{\nu\k}^{(gl)}(x,z) +\part_{x}^{\r}\r^{(gl)\nu\k}(x,z)\part_{z}^{\xi}F_{\nu\k}^{(gl)}(x,z) \) \right .  \nn\\
&& \left . \qquad-8 \( \part_{\mu}^{x}F^{(gl)\r\xi}(x,z)\part_{\b}^{z}\r^{(gl)\mu\b}(x,z)+\part_{\mu}^{x}\r^{(gl)\r\xi}(x,z)\part_{\b}^{z}F^{(gl)\mu\b}(x,z) \) \right . \nn\\
&& \left . \qquad+4 \( \part_{x}^{\r}F^{(gl)\nu\xi}(x,z)\part_{z}^{\b}\r_{\nu\b}^{(gl)}(x,z) +\part_{x}^{\r}\r^{(gl)\nu\xi}(x,z)\part_{z}^{\b}F_{\nu\b}^{(gl)}(x,z) \) \right . \nn\\
&& \left . \qquad-8 \(  \part_{\mu}^{x}F^{(gl)\mu\b}(x,z)\part_{\b}^{z}\r^{(gl)\r\xi}(x,z)+\part_{\mu}^{x}\r^{(gl)\mu\b}(x,z)\part_{\b}^{z}F^{(gl)\r\xi}(x,z) \) \right. \nn\\
&& \left . \qquad+4 \( \part_{x}^{\mu}F_{\mu\k}^{(gl)}(x,z)\part_{z}^{\xi}\r^{(gl)\r\k}(x,z) +\part_{x}^{\mu}\r_{\mu\k}^{(gl)}(x,z)\part_{z}^{\xi}F^{(gl)\r\k}(x,z) \) \right . \nn\\
&& \left . \qquad+4 \( \part_{\mu}^{x}F^{(gl)\mu\xi}(x,z)\part_{\b}^{z}\r^{(gl)\r\b}(x,z) +\part_{\mu}^{x}\r^{(gl)\mu\xi}(x,z)\part_{\b}^{z}F^{(gl)\r\b}(x,z) \) \right ]. \nn\\
&&\mbox{\ \ } 
\eea
Fermion propagator corrections are shown in Figure \ref{fig:fermion corr} and yield
\begin{figure}[hbt]
\centerline{ \epsfxsize 4in \epsfbox {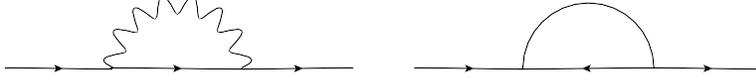} } \caption[]{
{\footnotesize Corrections to fermion propagator.}
} \label{fig:fermion corr}
\end{figure}
\bea
\Sigma_{F,\ \ldot\t}^{(f)}(x,z)&=&-2g^{2}N \left [ \s_{\a\ldot}^{\mu}\s_{\t\rdot}^{\nu} \( F^{(f)\a\rdot}(x,z)F_{\mu\nu}^{(gl)}(x,z) -\frac{1}{4}\r^{(f)\a\rdot}(x,z)\r_{\mu\nu}^{(gl)}(x,z) \) \right . \nn\\
&& \left . \qquad\ \ \ \  - 6 \( F_{\t\ldot}^{(f)}(z,x)F(x,z) + \frac{1}{4}\r_{\t\ldot}^{(f)}(z,x)\r(x,z) \) \right ], \\
&&\mbox{\ \ } \nn\\
\Sigma_{\r,\ \ldot\t}^{(f)}(x,z)&=&-2g^{2}N\left [ \s_{\a\ldot}^{\mu}\s_{\t\rdot}^{\nu} \( \r^{(f)\a\rdot}(x,z)F_{\mu\nu}^{(gl)}(x,z) + F^{(f)\a\rdot}(x,z)\r_{\mu\nu}^{(gl)}(x,z) \) \right . \nn\\
&& \left . \qquad\ \ \ \  + 6 \( \r_{\t\ldot}^{(f)}(z,x)F(x,z) - F_{\t\ldot}^{(f)}(z,x)\r(x,z) \) \right ].
\eea
Finally, the single ghost propagator correction is given in Figure \ref{fig:ghost corr},
\begin{figure}[hbt]
\centerline{ \epsfxsize 2in \epsfbox {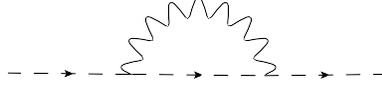} } \caption[]{
{\footnotesize Correction to ghost propagator.}
} \label{fig:ghost corr}
\end{figure}
with
\bea
\S_{F}^{(gh)}(x,z)&=& -g^{2}N\left [ 2\part_{x}^{\mu}F_{\mu\nu}^{(gl)}(x,z)\part_{z}^{\nu}F^{(gh)}(x,z) - \half\part_{x}^{\mu}\r_{\mu\nu}^{(gl)}(x,z)\part_{z}^{\nu}\r^{(gh)}(x,z) \right . \nn\\
&& \left . \qquad\ \ \  + 2 F_{\mu\nu}^{(gl)}(x,z)\part_{x}^{\mu}\part_{z}^{\nu}F^{(gh)}(x,z) - \half \r_{\mu\nu}^{(gl)}(x,z)\part_{x}^{\mu}\part_{z}^{\nu}\r^{(gh)}(x,z) \right ], \nn \\
&&\mbox{\ \ } \\
\S_{\r}^{(gh)}(x,z)&=& -2g^{2}N\left [ \part_{x}^{\mu}F_{\mu\nu}^{(gl)}(x,z)\part_{z}^{\nu}\r^{(gh)}(x,z) + \part_{x}^{\mu}\r_{\mu\nu}^{(gl)}\part_{z}^{\nu}F^{(gh)}(x,z) \right . \nn\\
&& \left . \qquad\ \ \  +  F_{\mu\nu}^{(gl)}(x,z)\part_{x}^{\mu}\part_{z}^{\nu}\r^{(gh)}(x,z) + \r_{\mu\nu}^{(gl)}(x,z)\part_{x}^{\mu}\part_{z}^{\nu}F^{(gh)}(x,z) \right ].
\eea

\bibliographystyle{JHEP}

\begin{thebibliography}{10}

\bibitem{Chesler:2008hg}
P.~M.~Chesler, L.~G.~Yaffe, {\it {Horizon formation and far-from-equilibrium isotropization in supersymmetric Yang-Mills plasma}}, {\em Phys.\ Rev.\ Lett.} {\bf 102} (2009) 211601, {{\tt arXiv:0812.2053}} [hep-th].

\bibitem{Maldacena:1997}
J.~M.~Maldacena, {\it {The large N limit of superconformal field theories and supergravity}}, {\em Adv.\  Theor.\ Math.\ Phys.} {\bf 2} (1998) 231-252, [{{\tt hep-th/9711200}}]. 

\bibitem{Aharony:1999}
O.~Aharony, S.~S.~Gubser, J.~M.~Maldacena, H.~Ooguri, and Y.~Oz, {\it {Large N field theories, string theory and gravity}}, {\em Phys.\ Rept.} {\bf 323} (2000) 183-386, [{{\tt hep-th/9905111}}].

\bibitem{Freedman:2002}
E.~DÕHoker and D.~Z.~Freedman, {\it {Supersymmetric gauge theories and the AdS/CFT correspondence}}, [{{\tt hep-th/0201253}}].

\bibitem{Nima:2010}
N.~Arkani-Hamed, F.~Cachazo and J.~Kaplan, {\it {What is the simplest quantum field theory?}}, {\em JHEP} 09:016, 2010, {{\tt arXiv:0808.1446}} [hep-th].


\bibitem{Hong:2006}
G.~Festuccia and H.~Liu, {\it {The arrow of time, black holes, and quantum mixing of large N Yang-Mills theories}}, {\em JHEP} 0712:027, 2007, [{{\tt hep-th/0611098}}].


 
\bibitem{Luttinger:1960}
J.~M.~Luttinger, J.~C.~Ward, {\it {Ground-state energy of a many-fermion system II}}, {\em Phys.\ Rev.} {\bf 118} (1960) 1417.

\bibitem{Baym:1962}
G.~Baym, {\it {Self-consistent approximations in many-body systems}}, {\em Phys.\ Rev.} {\bf127} (1962) 1391. 
 
\bibitem{Cornwall:1974}
J.~M.~Cornwall, R.~Jackiw and E.~Tomboulis, {\it {Effective Action For Composite Operators}}, {\em Phys.\ Rev.\ D} {\bf 10} (1974) 2428.

\bibitem{Smit:2002}
A.~Arrizabalaga and J.~Smit, {\it {Gauge-fixing dependence of Phi-derivable approximations}}, {\em Phys.\ Rev.\ D} {\bf 66} (2002) 065014, [{{\tt hep-ph/0207044}}].


\bibitem{Berges:2004_1}
J.~Berges, {\it {nPI effective action techniques for gauge theories}}, {\em Phys.\ Rev.\ D} {\bf 70} (2004) 105010, [{{\tt hep-ph/0401172}}].
 
\bibitem{Calzetta:1988}
E.~Calzetta and B.~L.~Hu, {\it {Nonequilibrium quantum fields: closed-time-path effective action, Wigner function, and Boltzmann equation}}, {\em Phys.\ Rev.\ D} {\bf 37} (1988) 2878.

\bibitem{Ivanov:1999}
Y. B. Ivanov, J. Knoll, and D. N. Voskresensky, {\it {Self-consistent approximations to non-equilibrium many-body theory}}, {\em Nucl.\ Phys.\ A} {\bf 657} (1999) 413-445, [{{\tt hep-ph/9807351}}].

\bibitem{Berges:2000}
J.~Berges and J.~Cox, {\it {Thermalization of quantum fields from time-reversal invariant evolution equations}}, {\em Phys.\ Lett.\ B} {\bf517} (2001) 369-374, [{{\tt hep-ph/0006160}}].

\bibitem{Aarts:2001}
G.~Aarts and J.~Berges, {\it {Nonequilibrium time evolution of the spectral function in quantum field theory}}, {\em Phys.\ Rev.\ D} {\bf 64} (2001) 105010.


\bibitem{Berges:2002}
J.~Berges, {\it {Controlled nonperturbative dynamics of quantum fields out of equilibrium}}, {\em Nucl.\  Phys.\ A} {\bf 699} (2002) 847.

\bibitem{Ahrensmeier:2002}
G.~Aarts, D.~Ahrensmeier, R.~Baier, J.~Berges and J.~Serreau, {\it {Far-from-equilibrium dynamics with broken symmetries from the 2PI-1/N expansion}},{\em Phys.\ Rev.\ D} {\bf 66} (2002) 045008, [{{\tt hep-ph/0201308}}].

\bibitem{Serreau:2002}
J.~Berges, Sz.~Bors‡nyi and J.~Serreau, {\it {Thermalization of fermionic quantum fields}}, {\em Nucl.\  Phys.\ B} {\bf 660} (2003) 52, [{{\tt hep-ph/0212404}}].

\bibitem{Juchem:2004}
S.~Juchem, W.~Cassing and C.~Greiner, {\it {Quantum dynamics and thermalization for out-of-equilibrium phi**4-theory}}, {\em Phys.\ Rev.\ D} {\bf 69} (2004) 025006.

\bibitem{Arriz:2004}
A.~Arrizabalaga, J.~Smit and A.~Tranberg, {\it {Tachyonic preheating using 2PI-1/N dynamics and the classical approximation}}, {\em JHEP} 0410:017, 2004, [{{\tt hep-ph/0409177}}].


\bibitem{Tranberg:2005}
A.~Arrizabalaga, J.~Smit and A.~Tranberg, {\it {Equilibration in phi**4 theory in 3+1 dimensions}}, {\em  Phys.\ Rev.\ D} {\bf 72} (2005) 025014, [{{\tt hep-ph/0503287}}].

\bibitem{Berges:2004}
J.~Berges, {\it {Introduction to nonequilibrium quantum field theory}}, {\em AIP\ Conf.\ Proc.} {\bf 739} (2005) 3-62, [{{\tt hep-ph/0409233}}]. 

\bibitem{Aarts:2003}
G.~Aarts and J.~M.~Martinez Resco, {\it{Transport coefficients from the 2PI effective action}}, {\em  Phys.\ Rev.\ D} {\bf 68} (2003) 085009, [{{\tt hep-ph/0303216}}].

\bibitem{Aarts:2004}
G.~Aarts and J.~M.~Martinez Resco, {\it{Shear Viscosity in the O(N) Model}}, {\em JHEP} 0402:061, 2004, [{{\tt hep-ph/0402192}}].

\bibitem{Aarts:2005}
G.~Aarts and J.~M.~Martinez Resco, {\it{Transport coefficients in Large $ N_{f} $ gauge theories with massive fermions}}, {\em JHEP} 0503:074, 2005, [{{\tt hep-ph/0503161}}].


\bibitem{Hu:2008}
E.~Calzetta and B.~Hu, {\it {Nonequilibrium quantum field theory}}, Cambridge University Press, 2008.

\bibitem{Moore:2007}
S.~C.~Huot, S.~Jeon and G.~D.~Moore, {\it{Shear viscosity in weakly coupled ${\cal N}=4$ Super Yang-Mills theory compared to QCD}}, {\em  Phys.\ Rev.\ Lett.} {\bf 98} (2007) 172303, [{{\tt hep-ph/0608062}}].

\bibitem{Balt:2008_1}
K.~Skenderis and B.~C.~van Rees, {\it {Real-time gauge/gravity duality}}, {\em Phys.\ Rev.\ Lett.} {\bf 101} (2008) 081601, {{\tt arXiv:0805.0150}} [hep-th].

\bibitem{Balt:2008_2}
K.~Skenderis and B.~C.~van Rees, {\it {Real-time gauge/gravity duality: prescription, renormalization and examples}}, {\em}, {{\tt arXiv:0812.2909}} [hep-th].

\bibitem{Balt:2009}
B.~C.~van Rees, {\it {Real-time gauge/gravity duality and ingoing boundary conditions}}, {\em Nucl.\ Phys.\ Proc.\ Suppl.} (2009)192-193:193-196, {{\tt arXiv:0902.4010}} [hep-th].

\bibitem{3loop1}
G.~D.~Moore, {\it{Transport coefficients at leading order: Kinetic theory versus diagrams}}, [{{\tt hep-ph/0211281}}]

\bibitem{3loop4}
M.~E.~Carrington, G.~Kunstatter and H.~Zakaret, {\it{2PI effective action and gauge dependence identities}}, {\em Eur.\ Phys.\ J\ C} {\bf 42} (2005) 253-259, [{{\tt hep-ph/0309084}}].

\bibitem{3loop7}
M.~E.~Carrington and E.~Kovalchuk, {\it{QED electrical conductivity using the two-particle-irreducible effective action}}, {\em  Phys.\ Rev.\ D} {\bf 76} (2007) 045019,{{\tt arXiv:0705.0162}} [hep-ph].

\bibitem{3loop8}
M.~E.~Carrington and E.~Kovalchuk, {\it{Leading order QED electrical conductivity from the three-particle irreducible effective action}}, {\em  Phys.\ Rev.\ D} {\bf 77} (2008) 025015, {{\tt arXiv:0709.0706}} [hep-ph].

\bibitem{3loop9}
U.~Reinosa and J.~Serreau, {\it{2PI functional techniques for gauge theories: QED}}, {\em  Annals\ of\ Physics} {\bf 325} (2010) 969-1017, {{\tt arXiv:0906.2881}} [hep-ph].

\bibitem{3loop10}
M.~E.~Carrington and E.~Kovalchuk, {\it{Leading order QCD shear viscosity from the three-particle irreducible effective action}}, {{\tt arXiv:0906.1140}} [hep-ph].

\bibitem{3loop11}
M.~E.~Carrington, {\it{Transport coefficients and nPI methods}}, {{\tt arXiv:1110.1238}} [hep-ph].

\bibitem{3loop12}
M.~E.~Carrington and E.~Kovalchuk, {\it{Towards next-to-leading order transport coefficients from the four-particle irreducible effective action}}, {\em  Phys.\ Rev.\ D} {\bf 81} (2010) 065017.

\bibitem{LPM1}
P.~Aurenche, F.~Gelis and H.~Zakaret, {\it{Landau-Pomeranchuk-Migdal effect in thermal field theory}}, {\em  Phys.\ Rev.\ D} {\bf 62} (2000) 096012.

\bibitem{LPM2}
P.~Aurenche, F.~Gelis, H.~Zakaret and R.~Kobes, {\it{Bremsstrahlung and photon production in thermal QCD}}, {\em  Phys.\ Rev.\ D} {\bf 58} (1998) 085003.

\bibitem{LPM3}
P.~Arnold, G.~D.~Moore and L.~G.~Yaffe, {\it{Photon emission from ultrarelativistic plasmas}}, [{{\tt hep-ph/0109064}}].

\bibitem{LPM4}
P.~Arnold, G.~D.~Moore and L.~G.~Yaffe, {\it{Photon and gluon emission in relativistic plasmas}}, [{{\tt hep-ph/0204343}}].

\bibitem{LPM5}
P.~Arnold, G.~D.~Moore and L.~G.~Yaffe, {\it{Effective kinetic theory for high temperature gauge theories}}, [{{\tt hep-ph/0209353}}].






\end{thebibliography}

\end{document}